\pdfoutput=1
\documentclass[sigconf,screen]{acmart}
\usepackage{booktabs} 
\usepackage[ruled]{algorithm2e} 
\usepackage{graphicx}
\usepackage{color, soul, xcolor}  
\usepackage{colortbl}
\usepackage{booktabs} 
\usepackage{amsmath}
\usepackage{enumitem}
\usepackage{balance}
\usepackage[normalem]{ulem}
\usepackage{caption}
\usepackage[caption=false,font=footnotesize,labelformat=simple]{subfig}
\usepackage{framed}
\definecolor{shadecolor}{gray}{0.95}

\acmConference[SWAN 2018]{The 4th International Workshop on Software Analytics (SWAN 2018)}{5 November, 2018}{Lake Buena Vista, Florida, USA}

\begin{document}

\copyrightyear{2018}
\acmYear{2018}
\setcopyright{acmcopyright}
\acmConference[SWAN '18]{Proceedings of the 4th ACM SIGSOFT International Workshop on Software Analytics}{November 5, 2018}{Lake Buena Vista, FL, USA}
\acmPrice{15.00}
\acmDOI{10.1145/3278142.3278149}
\acmISBN{978-1-4503-6056-2/18/11}

\title{Characterizing The Influence of Continuous Integration}\subtitle{Empirical Results from 250+ Open Source and Proprietary Projects}

\author{Akond Rahman, Amritanshu Agrawal, Rahul Krishna, and Alexander Sobran*}
\affiliation{North Carolina State University, IBM Corporation*} 
\email{[aarahman, aagrawa8, rkrish11]@ncsu.edu, asobran@us.ibm.com*}



\renewcommand{\shortauthors}{A. Rahman, A. Agrawal, R. Krishna, and A. Sobran}


\begin{abstract}

Continuous integration (CI) tools integrate code changes by automatically compiling, building, and executing test cases upon submission of code changes. Use of CI tools is getting increasingly popular, yet how proprietary projects reap the benefits of CI remains unknown. To investigate the influence of CI on software development, we analyze 150 open source software (OSS) projects, and 123 proprietary projects. For OSS projects, we observe the expected benefits after CI adoption, e.g., improvements in bug and issue resolution. However, for the proprietary projects, we cannot make similar observations. Our findings indicate that only adoption of CI might not be enough to the improve software development process. CI can be effective for software development if practitioners use CI's feedback mechanism efficiently, by applying the practice of making frequent commits. For our set of proprietary projects we observe practitioners commit less frequently, and hence not use CI effectively for obtaining feedback on the submitted code changes. Based on our findings we recommend industry practitioners to adopt the best practices of CI to reap the benefits of CI tools for example, making frequent commits.   

\end{abstract}

%
%


\maketitle

%

\section{Introduction}
\label{intro}

Continuous integration (CI) tools integrate code changes by automatically compiling, building, and executing test cases upon submission of code changes~\cite{Duvall:ci2007}. In recent years, usage of CI tools have become increasingly popular both for open source software (OSS) projects~\cite{beller:msr2017}~\cite{hilton:ase2016:ci}~\cite{vasilescu2015quality} as well as for proprietary projects~\cite{stahl:jss2017:ci:size}.

Our industrial partner adopted CI to improve their software development process. Our industrial partner's expectation was that similar to OSS projects~\cite{Zhao:ASE17:CI}~\cite{vasilescu2015quality}, CI  would positively influence resolution of bugs and issues for projects owned by our industrial partner. Our industrial partner also expected collaboration to increase upon adoption of CI. Being one of the primary Extreme Programming (XP) practices~\cite{Beck:1999:XP}, CI is expected to benefit collaboration amongst team members~\cite{Sharp:2008:XP:colla}. 

We conduct an empirical study to investigate if our industrial partner's expectations were fulfilled. Such an empirical study can be beneficial in the following ways: (i) to quantify if CI benefits projects with respect to bug and issue resolution, along with collaboration; and (ii) to derive lessons that industry practitioners should keep in mind when using CI. We conduct an empirical study with 150 OSS and 123 proprietary projects to quantify the influence of CI on bug resolution, collaboration, and issue resolution. We answer the following research questions: 

\begin{description}[leftmargin=*]
\item{\textbf{\textit{RQ1: Does adoption of continuous integration influence commit patterns?}}} Commit frequency and sizes significantly increases for OSS projects after CI adoption but not for our set of proprietary projects. 

\item{\textbf{\textit{RQ2: How does adoption of continuous integration influence collaboration amongst team members?}}} After adopting CI, collaboration significantly increases for both, OSS and our set of proprietary projects. The increase in collaboration is more observable for OSS projects than the proprietary projects. 

\item{\textbf{\textit{RQ3: How does adoption of continuous integration influence bug and issue resolution?}}} Significantly more bugs and issues are resolved after adoption of CI for OSS projects, but not for our set of proprietary projects.

\end{description}

In summary, we observe usage of CI to be beneficial for OSS projects but not for our set of proprietary projects. For proprietary projects, we acknowledge that there may be benefits to CI  which are not captured by our study, for example, cultural benefits in adopting CI tools. Findings from our paper can help industry practitioners revise their expectations about the benefits of CI. Our paper may also help to identify possible strategies to fully reap the benefits of CI.  




\begin{figure*}
\includegraphics[width=0.95\textwidth]{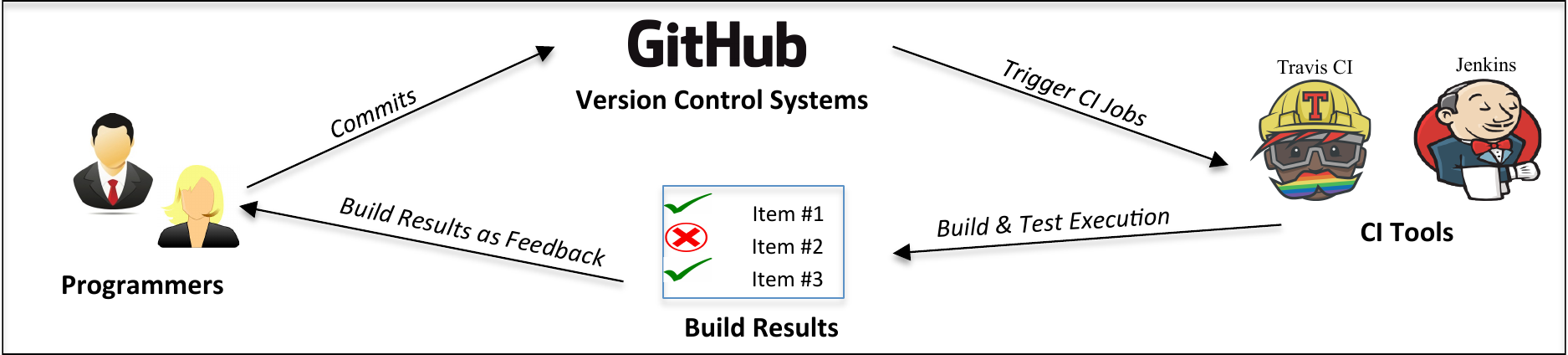}
\setlength\abovecaptionskip{-1pt}
\setlength\belowcaptionskip{-1pt}
\caption{An example work-flow of the continuous integration (CI) process.}
\label{figure-ci-bg}
\end{figure*}

\section{Background}
\label{bg:rel}
We first provide a brief background on CI, then describe prior research work related to CI.

\subsection{About Continuous Integration (CI)}
CI is identified as one of the primary practices to implement XP~\cite{Beck:1999:XP}. According to Duvall et al.~\cite{Duvall:ci2007}, CI originated from the imperatives of agility, in order to respond to customer requests quickly. When building the source code, CI tools can execute unit and integration tests to ensure quality of the integrated source code. If the tests do not pass, CI tools can be customized to give feedback on the submitted code changes. Even though the concept of CI was introduced in 2006, initial usage of CI was not popular amongst practitioners~\cite{Deshpande2008:ci}. However, since 2011, with the advent of CI tools such as Travis CI~\cite{travis:ci}, usage of CI has increased in recent years~\cite{hilton:ase2016:ci}.

When a software team adopts CI, the team has to follow a set of practices~\cite{Duvall:ci2007}. According to the CI methodology all programmers have to check-in their code daily, which are integrated daily~\cite{Duvall:ci2007}. Unlike, traditional methodologies such as waterfall, in CI, programmers get instant feedback on their code via build results. To implement CI, the team must maintain its source code in a version control system (VCS), and integrate the VCS with the CI tool so that builds are triggered upon submission of each commit~\cite{Duvall:ci2007}. Figure~\ref{figure-ci-bg} provides an example on how a typical CI process works. Programmer make commits in a repository maintained by a VCS such as, GitHub, and these commits trigger CI jobs on a CI tool such as Travis CI which executes, builds, tests, and produces build results. These build results are provided to the programmers as a feedback either through e-mails, or phone alerts~\cite{Duvall:ci2007} on their submitted code changes. Based on the build results, programmers make necessary changes to their code, and repeats the CI process again.

\subsection{Related Work}

Our paper is closely related to prior research that have investigated usage of CI tools. We briefly describe these prior work as following. 

\textbf{Adoption}: Hilton et al.~\cite{hilton:ase2016:ci} mined OSS projects hosted on Github. They observed that most popular projects use CI, and reported that the median time of CI adoption is one year. They also advocated for wide-spread adoption of CI, as CI correlates with several positive outcomes. However, adoption of CI is non-trivial as suggested by other prior work; for example, Olsson et al.~\cite{olsson:stairway:2012} who identified lack of automated testing frameworks as a key barrier to transition from a traditional software process to a CI-based software process. Also, Hilton et al.~\cite{Hilton:fse2017:ci} surveyed industrial practitioners and identified three trade-offs to adopt CI: assurance, flexibility, and security. Rahman et al.~\cite{Rahman:RCOSE17} observed that adoption of CI is not wide-spread amongst practitioners. They investigated which diffusion of innovation (DOI) factors influence adoption of CI tools, and reported four factors: relative advantages, compatibility, complexity, and education. 

\textbf{Usage}: Beller et al.~\cite{beller:msr2017} collected and analyzed Java and Ruby-based projects from Github, and synthesized the nature of build and test attributes exhibited amongst OSS projects that use CI. Vasilescu et al.~\cite{vasilescu2015quality} analyzed OSS GitHub projects that use Travis CI, and reported that adoption of CI increases productivity for OSS projects. Zhao et al.~\cite{Zhao:ASE17:CI} mined OSS GitHub projects, and investigated if software development practices such as commit frequency, commit size, and pull request handling, changes after adoption of CI.    

The above-mentioned findings highlight the community's interest in how CI is being used in software projects. From the above-mentioned prior work, we can list the following as exemplars of the expected benefits of adopting CI:
\begin{itemize}[leftmargin=*]
\item Zhao et al.~\cite{Zhao:ASE17:CI} reported that for OSS GitHub projects, the number of closed issues increases after the adoption of CI tools.
\item Vasilescu et al.~\cite{vasilescu2015quality} reported that for OSS GitHub projects, number of bugs do not increase after adoption of CI.
\end{itemize}

Note that all of these findings are derived from OSS projects. With respect to the development process, structure, and complexity, proprietary projects are different from OSS projects~\cite{Paulson:TSE2004}~\cite{Robinson:esem2010:prop:oss}, which motivates us to pursue our research study. Hence, for the rest of this paper, we will compare the influence of adopting CI within OSS and our set of proprietary projects. We consider the following attributes of software development: bug resolution, collaboration amongst team members, commit patterns, and issue resolution.


\section{Methodology}
\label{meth}

In this section, we describe our methodology to filter datasets, followed by metrics and statistical measurements that we use to answer our research questions.  
 
\subsection{Filtering}
\label{meth:filer}

We conduct our empirical study using OSS projects from GitHub, and proprietary projects collected from our industrial partner. For OSS projects we selected public GitHub projects that are included as a `GitHub showcase project'. Of the publicly available projects hosted on GitHub, a selected set of projects are marked as `showcases', to demonstrate how a project can be developed in certain domain~\cite{gh:showcase}. Example projects include: Javascript libraries such as `AngularJS'\footnote{https://github.com/angular/angular.js}, and programming languages such as `Scala'\footnote{https://github.com/scala/scala}. Our assumption is that by selecting these GitHub projects we can start with a representative set of OSS projects that enjoy popularity, and provide good examples of software development. We extracted the data from OSS and proprietary projects by using the public GitHub API and a private API maintained by our industrial partner respectively. 


Projects that are hosted on GitHub provides researchers the opportunity to extract necessary project information such as commits, and issues~\cite{kalliamvakou2014promises}~\cite{bird2009promises}. Unfortunately, these projects can contain short development activity, and not be related to software development at all~\cite{kalliamvakou2014promises}~\cite{bird2009promises}. Hence, we need to curate a set of projects that can contain sufficient software development data for analysis. We apply a filtering strategy that can be described in the following manner: 

\begin{itemize}[leftmargin=*]
\item{
\textbf{Filter-1 (General):}
As the first step of filtering, we identify projects that contain sufficient software development information using the criteria used by prior research~\cite{amrit:icse:seip2018}~\cite{rahul:icse18:seip}. By applying these filtering criteria we mitigate the limitations of mining GitHub projects stated by prior researchers~\cite{kalliamvakou2014promises}~\cite{bird2009promises}.
}
\item{
\textbf{Filter-2 (CI)}
We use the second filter to identify projects that have adopted CI tools.  
\begin{itemize}[leftmargin=*]
\item{\textit{CI Tool Usage}: The project must use any one of the following tools: Circle CI, Jenkins, and Travis CI. We select these tools as these tools are frequently used in GitHub projects~\cite{hilton:ase2016:ci}. We determine if a project is using Circle CI, Jenkins, and Travis CI by inspecting the existence of `circle.yml', `jenkins.yml', and `travis.yml', respectively, in the root directory of the project. }
\item{\textit{Start Date}: The project must start on or after January, 2014. From our initial exploration we observe that 90\% of the collected proprietary projects start on or after 2014. }
\end{itemize}
}
\end{itemize}


\subsection{Metrics}
\label{meth-metric}

We use the metrics presented in Table~\ref{table-meth-metrics} to answer our research questions. The `Metric Name' column presents the metrics, and the `Equation' presents the corresponding equation for each metric.   

\begin{table*}[]
\centering
\setlength\abovecaptionskip{-1pt}
\caption{Metrics Used to Answer RQ1, RQ2, and RQ3}
\label{table-meth-metrics}
{\footnotesize
\begin{tabular}{ p{4.8cm} | p{2cm} | p{3.8cm}  }
\hline
\textbf{Metric Name} & \textbf{Equation}    & \textbf{Brief Description}  \\
\hline
Proportion of Closed Issues ($CLI$) 
& \begin{equation}
CLI(p, m) = \frac{\text{total count of closed issues in month $m$, for project $p$}}{\text{total count of issues in month $m$, for project $p$}}\label{equ:ci}
\end{equation}             
& Count of closed issues per month  \\
\hline
Normalized Proportion of Closed Issues ($NCI$)
& \begin{equation}
NCI (p) =  \dfrac{\sum_{i=1}^{M}CLI(p, i)}{M}\label{equ:nci}
\end{equation}           
& $CLI$ normalized by time  \\
\hline
Proportion of Closed Bugs ($CB$) 
& \begin{equation}
CB(p, m) = \frac{\text{total count of closed bugs in month $m$, for project $p$}}{\text{total count of bugs in month $m$, for project $p$}}\label{equ:cb}
\end{equation} 
& Count of closed bugs per month \\
\hline
Normalized Proportion of Closed Bugs ($NCB$) 
& \begin{equation}
NCB (p) =  \dfrac{\sum_{i=1}^{M}CB(p, i)}{M}\label{equ:ncb}
\end{equation}
& $CB$ normalized by time  \\
\hline
Count of Non-Merge Commits ($CC$) 
& 
\begin{equation}
CC(p, m) = \frac{\text{total count of non-merge commits in month $m$, for project $p$}}{\text{total count of active programmers in month $m$, for project $p$}}\label{equ:cc}
\end{equation}
& Count of non-merge commits per month, normalized by the number of programmers\\
\hline
Normalized Count of Commits ($NCC$) 
& 
\begin{equation}
NCC (p) =  \dfrac{\sum_{i=1}^{M}CC(p, i)}{M}\label{equ:ncc}
\end{equation}
& $CC$ normalized by time  \\
\hline
Commit size ($CS$)
& 
\begin{equation}
CS(p, m) = \frac{\text{total lines added and deleted in month $m$, for project $p$}}{\text{total count of commits in month $m$, for project $p$}}\label{equ:cs}
\end{equation}
& Total lines of code added and deleted per commit within a month   \\
\hline
Normalized Commit Size ($NCS$)
& 
\begin{equation}
NCS (p) =  \dfrac{\sum_{i=1}^{M}CS(p, i)}{M}\label{equ:ncs}
\end{equation}
& $CS$ normalized by time   \\
\hline
Median In-degree ($MID$)
& 
\begin{equation}
\text{Median In-Degree (MID)} = \frac{\text{median in-degree}}{\text{total count of nodes}}\label{equ:mid}
\end{equation}
& In-degree corresponds to collaboration between the programmers. The higher the median in-degree, the higher connection is between the nodes~\cite{Bhattacharya:ICSE2012:GAP}, indicating more collaboration between the programmers.   \\
\hline
Normalized Median In-degree ($NMID$)
& 
\begin{equation}
\text{Normalized Median In-Degree (NMID)} =  \dfrac{\sum_{i=1}^{M}MID(i)}{M}\label{equ:nmid}
\end{equation}
& $MID$ normalized by time   \\
\hline
\end{tabular}
}
\end{table*}

According to Table~\ref{table-meth-metrics}, the metrics Normalized Proportion of Closed Issues ($NCI$), Normalized Proportion of Closed Bugs ($NCB$), Normalized Count of Commits ($NCC$), Normalized Commit Size ($NCS$), and Normalized Median In-degree ($NMID$) are normalized by $M$. Here, $M$ presents the count of months before or after adoption of CI for a project. For example, if the number of months before and after adoption of CI is respectively, 20 and 30 then, we use Equation~\ref{equ:nci} with $M=20$ to calculate the project's $NCI$ before adoption of CI, and with $M=30$, to calculate the project's $NCI$ after adoption of CI. In a similar fashion, we calculate $NCB$, $NCC$, $NCS$, and $NMID$ by using $M$ i.e., months before or after adoption of CI. 



\begin{figure}
\subfloat[]{
 \includegraphics[width=0.15\textwidth]{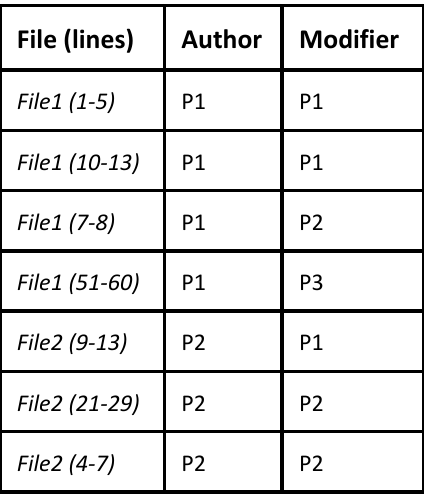}
 \label{fig:meth:graph1}
}
\subfloat[]{
 \includegraphics[width=0.15\textwidth]{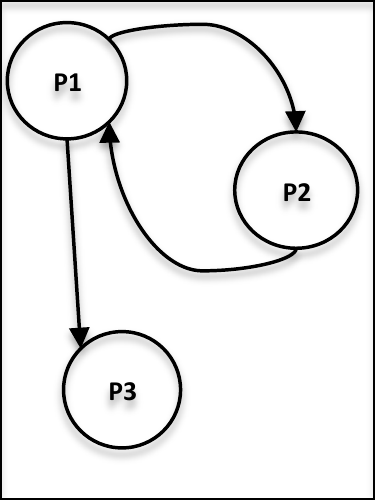}
 \label{fig:meth:graph2}
}
\subfloat[]{
 \includegraphics[width=0.15\textwidth]{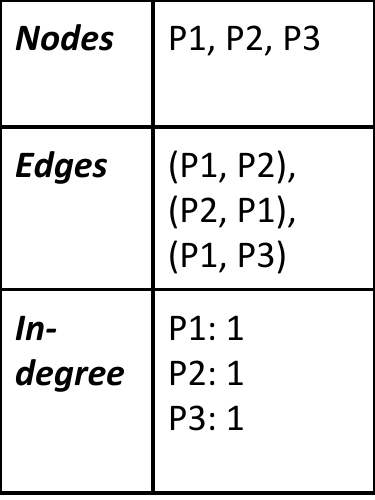}
 \label{fig:meth:graph3}
}
\setlength\abovecaptionskip{-2pt}
\setlength\belowcaptionskip{-2pt}
\caption{Hypothetical example on how we construct collaboration graphs.} 
\label{fig:meth:graph} 
\end{figure}

Figure~\ref{fig:meth:graph} provides a hypothetical example to calculate metric `Median In Degree'. We observe a list of programmers who are authoring and modifying two files. We construct a graph, using the modification information, as shown in Figure~\ref{fig:meth:graph2}. The constructed graph has three nodes (P1, P2, and P3), and three edges. In our hypothetical example, the project's collaboration graph has three edges, and the in-degree for nodes P1, P2, and P3 is one. Therefore, the median in-degree for the collaboration graph is one. 

\subsection{Statistical Measurements}
\label{meth:stat}

We use three statistical measures to compare the metrics of interest before and after adoption of CI: effect size using Cliff's Delta~\cite{cliff1993:original}, the Mann-Whitney U test~\cite{mann:whitney:original}, and the `delta ($\Delta$)' measure. Both, Mann-Whitney U test and Cliff's Delta are non-parametric. The Mann-Whitney U test states if one distribution is significantly large/smaller than the other, whereas effect size using Cliff's Delta measures how large the difference is. Following convention, we report a distribution to be significantly larger than the other if $p-value <0.05$. We use Romano et al.'s recommendations to interpret the observed Cliff's Delta values. According to Romano et al.~\cite{Romano:CliffsCutoff2006}, the difference between two groups is `large' if Cliff's Delta is greater than 0.47. A Cliff's Delta value between 0.33 and 0.47 indicates a `medium' difference. A Cliff's Delta value between 0.14 and 0.33 indicates a `small' difference. Finally, a Cliff's Delta value less than 0.14 indicates a `negligible' difference. 

We also report `delta ($\Delta$)', which is the difference between the median values, before and after adoption of CI. The `delta' measurement quantifies the proportion of increase or decrease, after and before adoption of CI. As a hypothetical example, for OSS projects, if median $NCI$ is 10.0, and 8.5, respectively, after and before adoption of CI, then the `delta ($\Delta$)' is +0.17 (= (10-8.5)/8.5).

\section{Results}
\label{results}

Before providing the answers to the research questions, we present summary statistics of the studied projects. Initially we started with 1,108 OSS projects and 538 proprietary projects. Upon applying Filter-1 we are left with 661 open source and 171 proprietary projects. As shown in Table~\ref{res:table:sanity2}, after applying Filter-2, we are finally left with 150 OSS and 123 proprietary projects. We use these projects to answer the three research questions. A brief summary of the filtered projects is presented in Table~\ref{table-summary}. The commit count per programmer is 24.2 and 46.7, respectively for OSS and proprietary projects. On average a programmer changes 141 and 345 files, respectively for OSS and proprietary projects.

\begin{table}
\footnotesize
\centering
\setlength\abovecaptionskip{-1pt}
\caption{Projects filtered for each sanity check of Filter-2.}
\begin{tabular}{|l|r|r|}
\hline
Sanity check                               & OSS  & Proprietary \\
\hline
CI Tool Usage                              & 448  & 46          \\
Start Date (Must start on or after 2014)   & 63   & 2           \\
\hline
Project count after filtering              & 150 & 123          \\  
\hline
\end{tabular}
\label{res:table:sanity2}
\end{table}

\begin{table}[]
\centering
\setlength\abovecaptionskip{-1pt}
\caption{Summary of Projects}
\label{table-summary}
{\footnotesize
\begin{tabular}{ p{3.5cm} p{2cm} p{2cm}  }
\hline
\textbf{Property}   & \textbf{Project Type} \\
                    & OSS                   & Proprietary  \\
\hline
Total Changed Files & 1,122,352             & 728,733 \\
Total Commits       & 191,804               & 98,542 \\
Total LOC Added     & 48,424,888            & 44,003,385 \\
Total LOC Deleted   & 30,225,543            & 26,614,230 \\
Total Programmers   & 7,922                 & 2,109 \\
\hline
\textbf{Total Projects}      & 150                   & 123 \\
\hline
\end{tabular}
}
\end{table}

\subsection{Answer to RQ1: Does adoption of continuous integration influence commit patterns?}
\label{res:rq4}

\begin{figure*}
\subfloat[]{
 \includegraphics[width=0.24\textwidth]{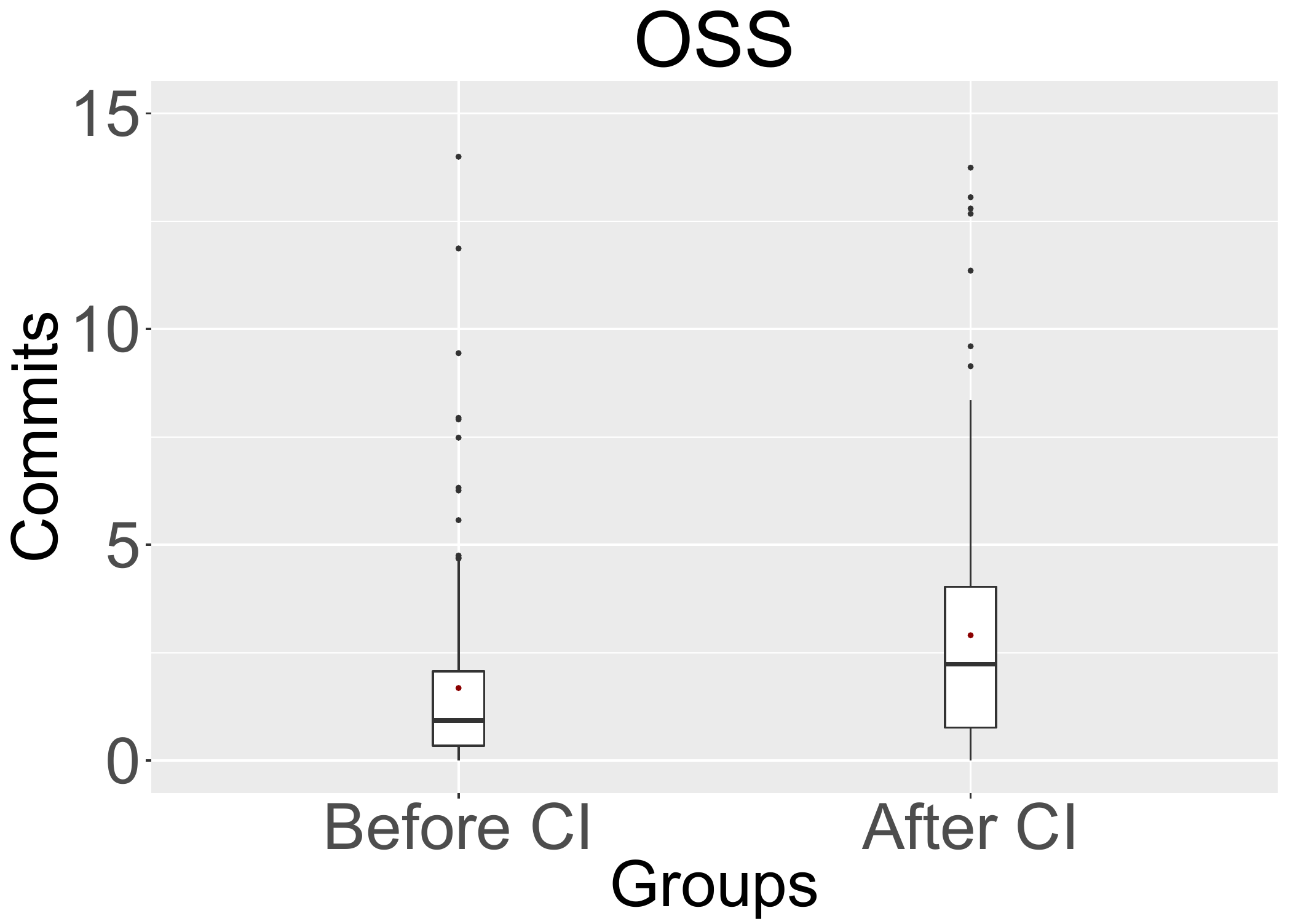}
 \label{fig:res:ncc:oss}
}
\subfloat[]{
 \includegraphics[width=0.24\textwidth]{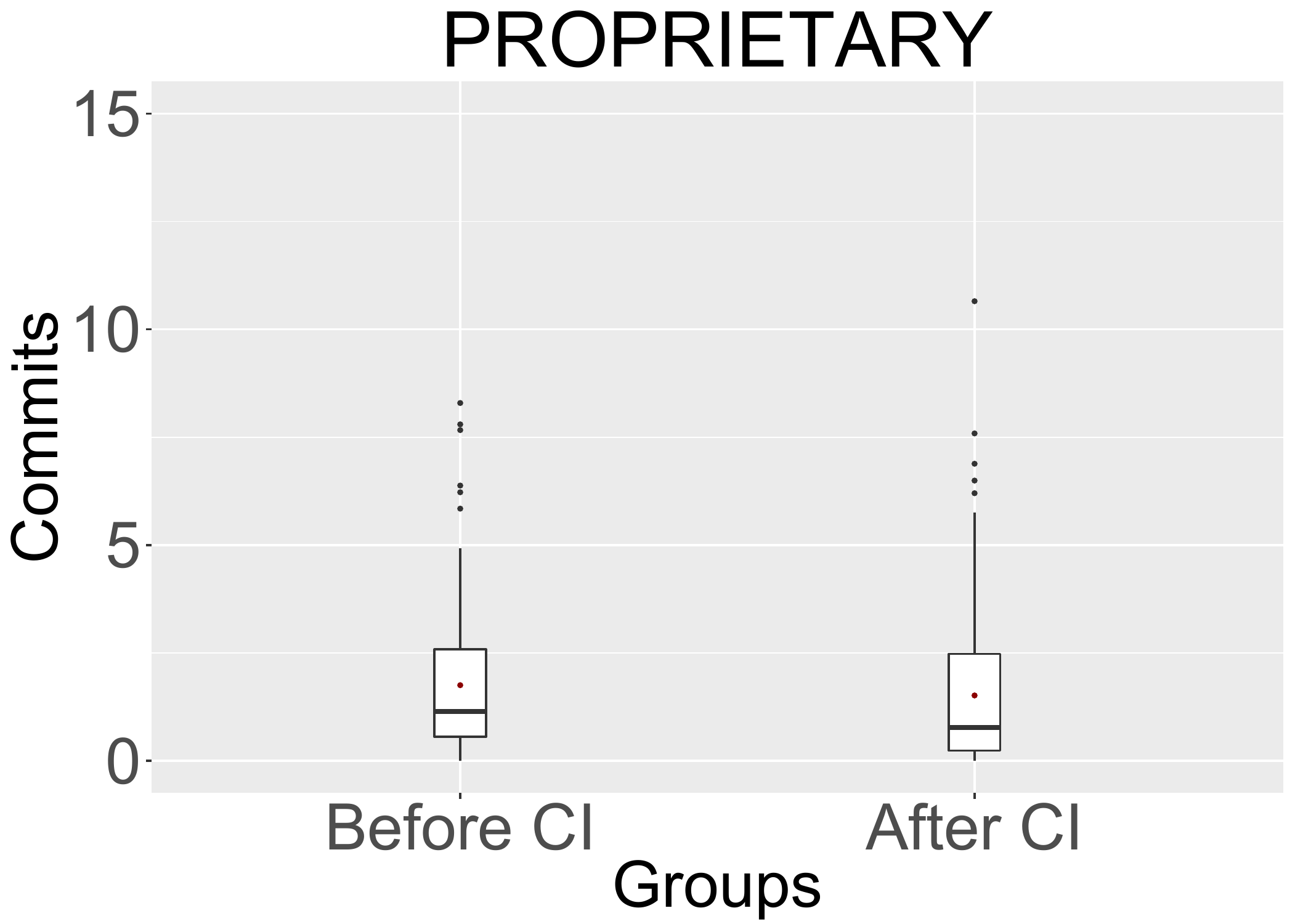}
 \label{fig:res:ncc:pro}
}
\subfloat[]{
 \includegraphics[width=0.24\textwidth]{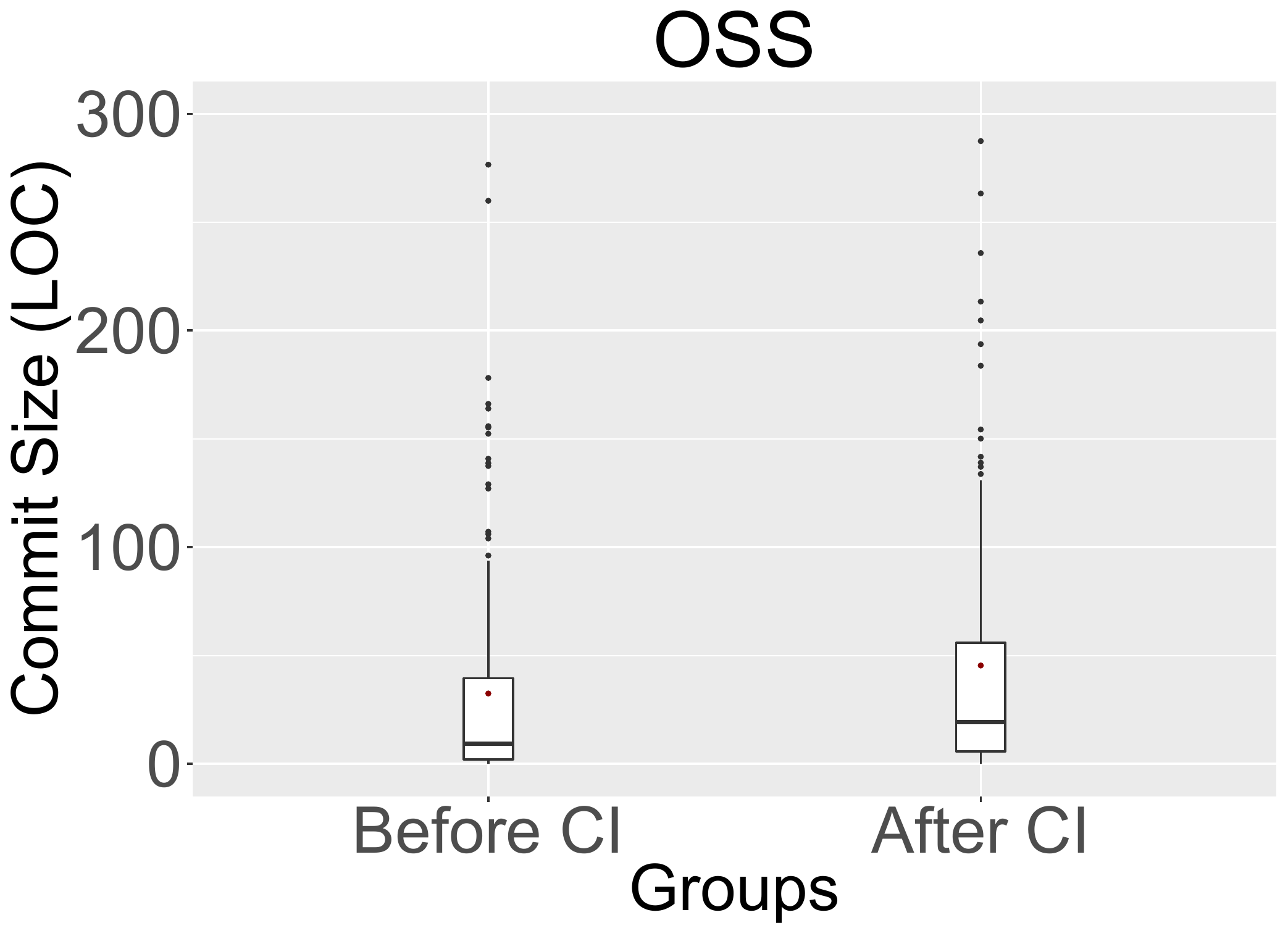}
 \label{fig:res:ncs:oss}
}
\subfloat[]{
 \includegraphics[width=0.24\textwidth]{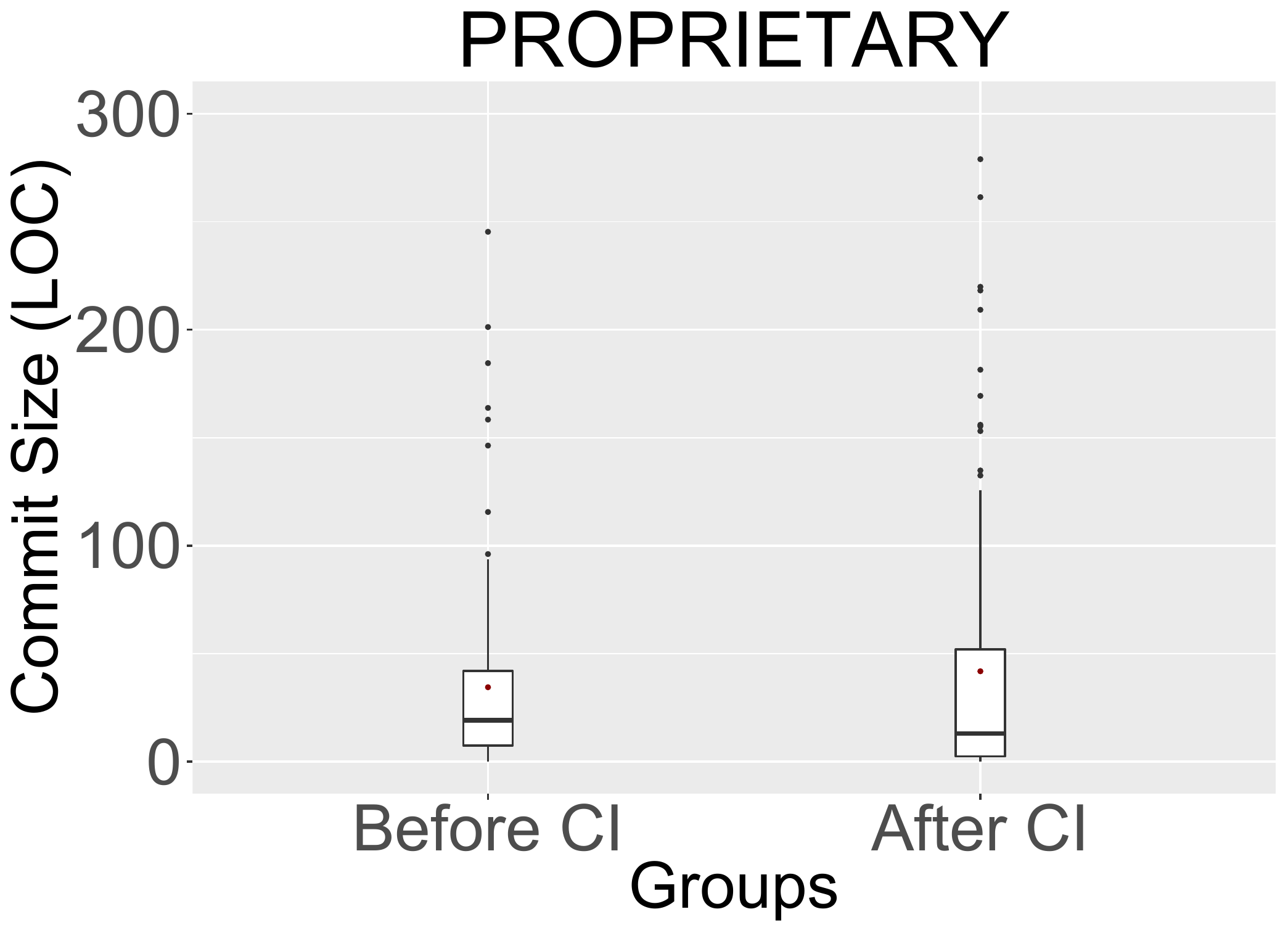}
 \label{fig:res:ncs:pro}
}
\setlength\abovecaptionskip{-2pt}
\setlength\belowcaptionskip{-2pt}
\caption{Normalized commit count ($NCC$) and normalized commit size ($NCS$) for our set of OSS and proprietary projects. Figures~\ref{fig:res:ncc:oss} and~\ref{fig:res:ncc:pro} respectively presents the normalized count of commits in OSS and proprietary projects. Figures~\ref{fig:res:ncs:oss} and ~\ref{fig:res:ncs:pro} respectively presents the normalized commit sizes for OSS and proprietary projects.} 
\label{fig:res:rq4} 
\end{figure*}

Zhao et al.~\cite{Zhao:ASE17:CI} mined OSS GitHub projects, and reported that after adoption of CI, frequency of commits increases. We expect that our answers to RQ1 for OSS projects will be consistent with Zhao et al.'s~\cite{Zhao:ASE17:CI} findings. We answer RQ1, by first reporting the frequency of commits before and after adoption of CI. We report the results of the three statistical measures in Table~\ref{table:rq4} and the box-plots in Figure~\ref{fig:res:rq4}. The `delta' metric is represented in the $\Delta$ row. The `delta' value for which we observe no significant difference is highlighted in grey. 

Our findings indicate that for proprietary projects, programmers are not making frequent commits after adoption of CI. On the contrary for OSS projects programmers are making significantly more commits, confirming findings from prior research~\cite{Zhao:ASE17:CI}.

Commit size is another measure we use to answer RQ1. As shown in Table~\ref{table:rq4} we observe size of commits i.e., churned lines of code per commit to significantly increase for OSS projects, but not for proprietary projects. 

\noindent
\fbox{\begin{minipage}{26em}
\textbf{\textit{Answer to RQ1: After adoption of CI, normalized commit frequency and commit size significantly increases for our set of OSS projects, but not for our set of proprietary projects. For proprietary projects we do not observe CI to have an influence on normalized commit frequency and commit size.}}
\end{minipage}}

\begin{table}
\centering
\setlength\abovecaptionskip{-2pt}
\caption{Influence of CI on Commit Patterns.}
\footnotesize{
\begin{tabular}{ |p{1cm}|p{1.25cm}|p{1.25cm}|p{1.25cm}|p{1.25cm}| } 
\hline
            & \textbf{Commit Count} ($NCC$) &          & \textbf{Commit Size} ($NCS$) &  \\ 
\hline
Measure     & OSS             & Prop.          & OSS                   & Prop. \\ 
\hline
Median      & (A:2.2, B:0.9)  & (A:0.7, B:1.1) & (A:25.2, B:10.5)     & (A:14.6, B:23.8)   \\ 
$\Delta$    & +1.44           & \cellcolor{lightgray}-0.36          & +1.40                & \cellcolor{lightgray}-0.38              \\ 
p-value     & $< 0.001$       & 0.9            & 0.001                & 0.9                 \\ 
Effect size & 0.3             & 0.1            & 0.2                  & 0.1                 \\ 
\hline
\end{tabular}
}
\label{table:rq4}
\end{table}


\subsection{Answer to RQ2: How does adoption of continuous integration influence collaboration amongst team members?}
\label{res:rq3}

\begin{figure}
\subfloat[]{
 \includegraphics[width=0.24\textwidth]{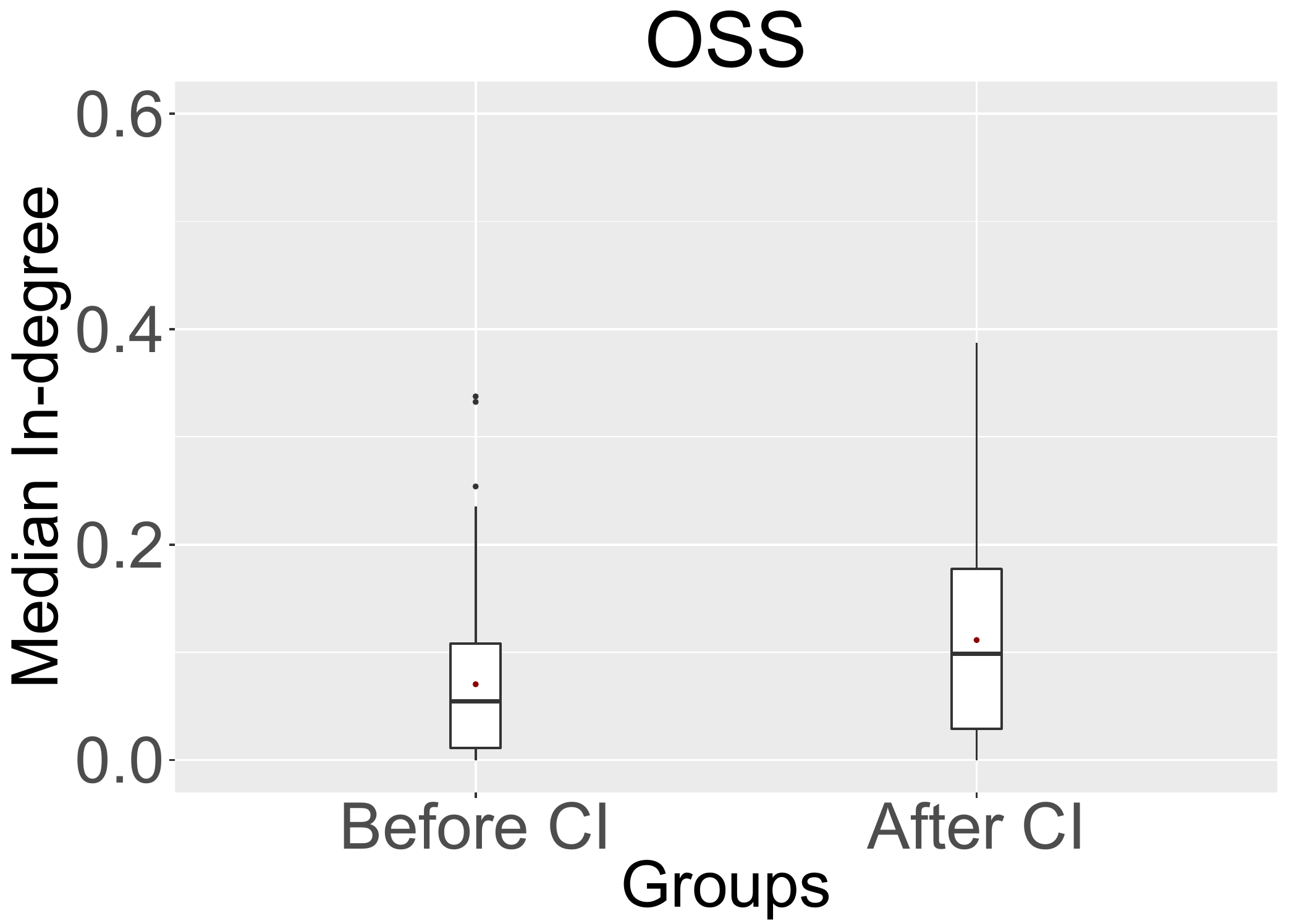}
 \label{fig:res:nmid:oss}
}
\subfloat[]{
 \includegraphics[width=0.24\textwidth]{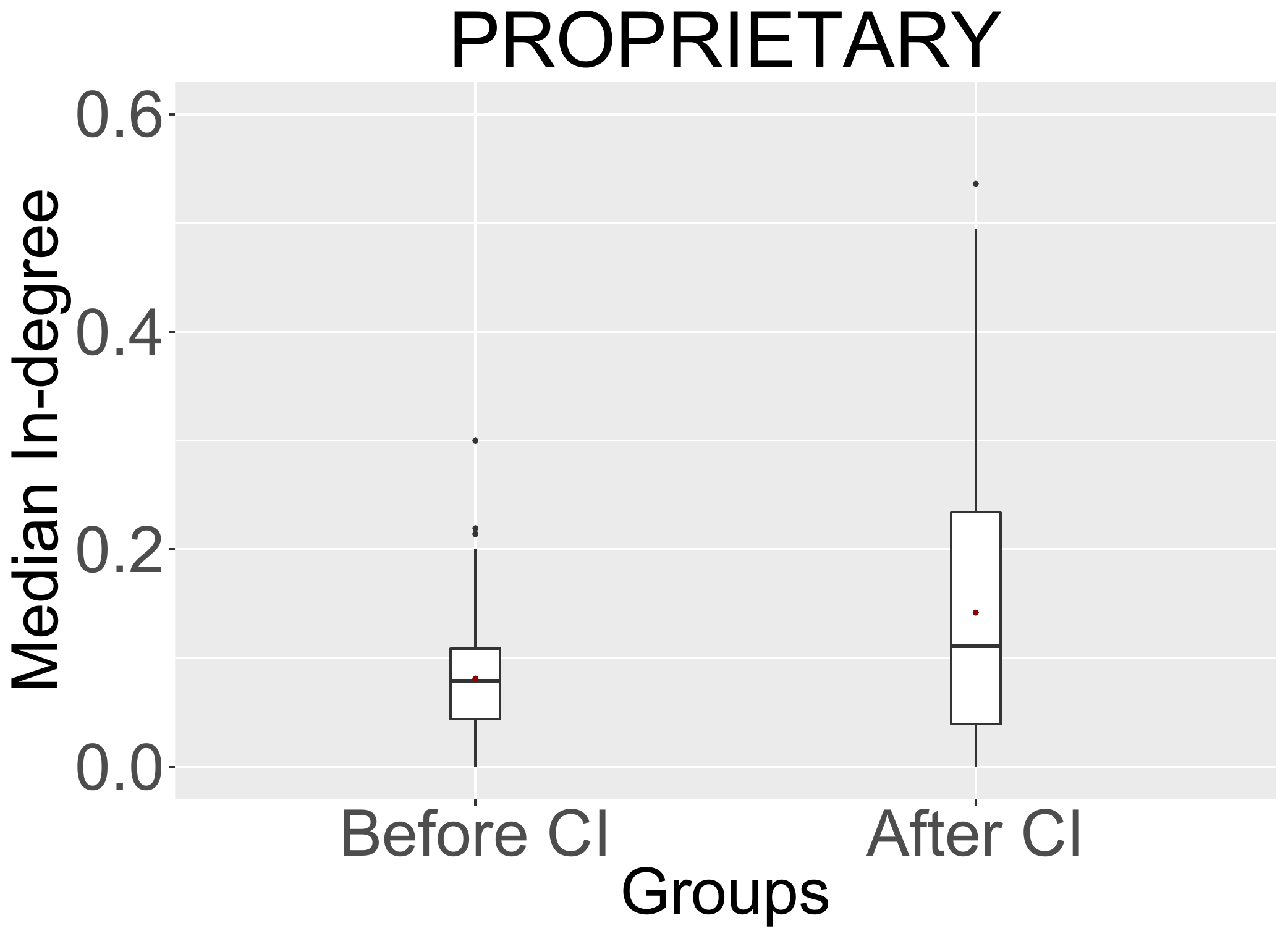}
 \label{fig:res:nmid:pro}
}
\setlength\abovecaptionskip{-3pt}
\setlength\belowcaptionskip{-3pt}
\caption{Influence of CI on collaboration ($NMID$).} 
\label{fig:res:rq3} 
\end{figure}

As described in Section~\ref{meth-metric}, we report the normalized median in-degree (NMID) to answer RQ2. We report the summary statistics in Table~\ref{table:rq3}, and the box-plots in Figure~\ref{fig:res:rq3}. For both OSS and proprietary projects, the median in-degree significantly increases after adoption of CI. The effect size for OSS and proprietary projects is 0.2, which is small according to Romano et al~\cite{Romano:CliffsCutoff2006}. Based on the `delta' measure ($\Delta$ in Table~\ref{table:rq3}) we observe that the increase in collaboration is not as high for proprietary projects, as it is for OSS projects. 

\noindent
\fbox{\begin{minipage}{26em}
\textbf{\textit{Answer to RQ2: After adoption of CI, normalized collaboration amount between programmers significantly increases for our set of OSS and proprietary projects. That said, increase in collaboration is larger for OSS projects, compared to proprietary projects.}}
\end{minipage}}

\begin{table}
\centering
\setlength\abovecaptionskip{-2pt}
\caption{Influence of CI on Collaboration ($NMID$)}
\footnotesize{
\begin{tabular}{ |c|c|c| } 
\hline
Measure     & OSS           & Prop.                        \\ 
\hline
Median      & (A: 0.09, B:0.05)      & (A: 0.11, B:0.07)    \\ 
$\Delta$    & +0.8                   & +0.5                 \\ 
p-value     & $< 0.001$     & $< 0.001$                      \\ 
Effect size & 0.2           & 0.2                            \\ 
\hline
\end{tabular}
}
\label{table:rq3}
\end{table}

\subsection{Answer to RQ3: How does adoption of continuous integration influence bug and issue resolution?}
\label{res:rq1}

\begin{figure*}
\subfloat[]{
 \includegraphics[width=0.24\textwidth]{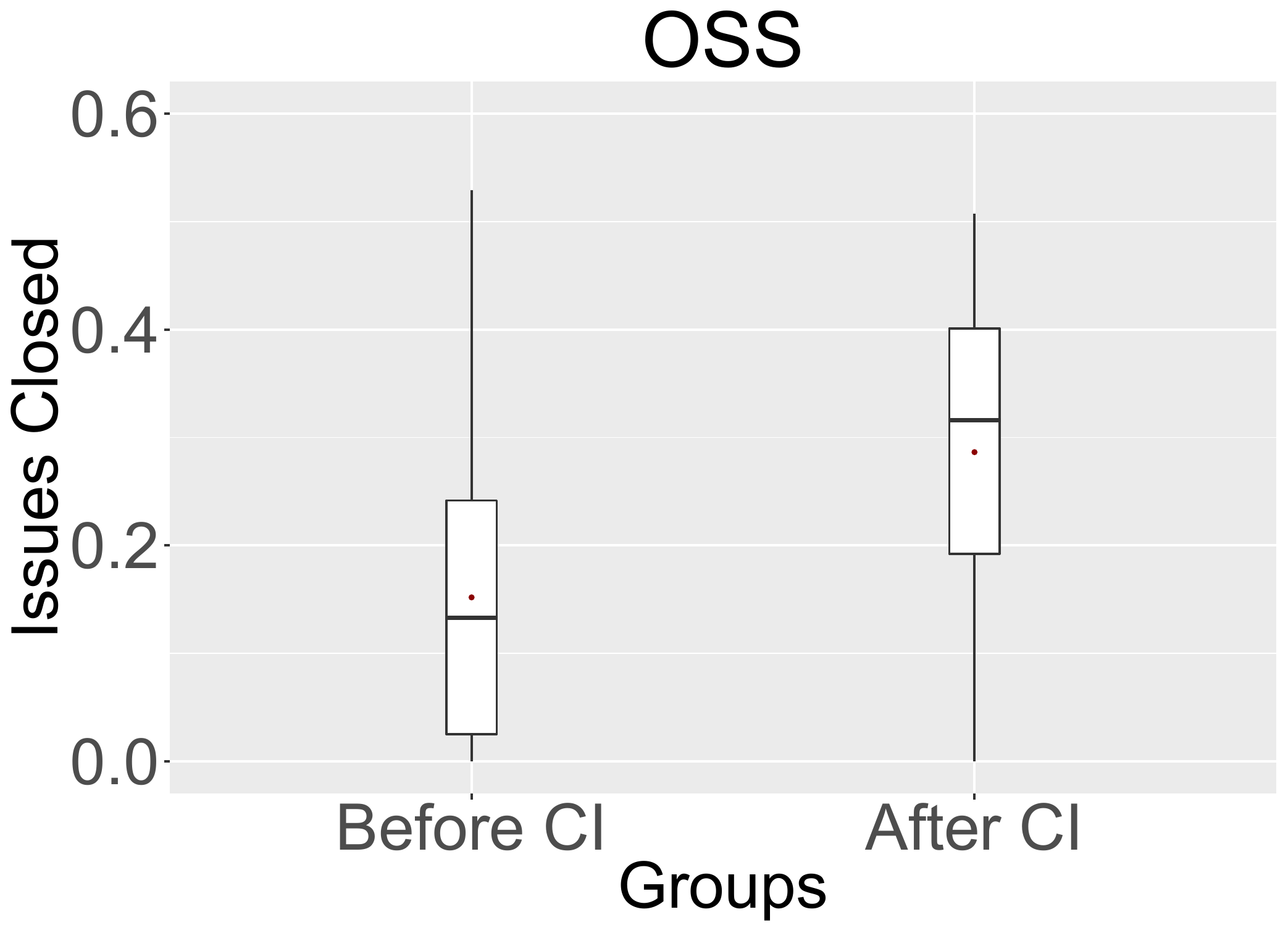}
 \label{fig:res:nci:oss}
}
\subfloat[]{
 \includegraphics[width=0.24\textwidth]{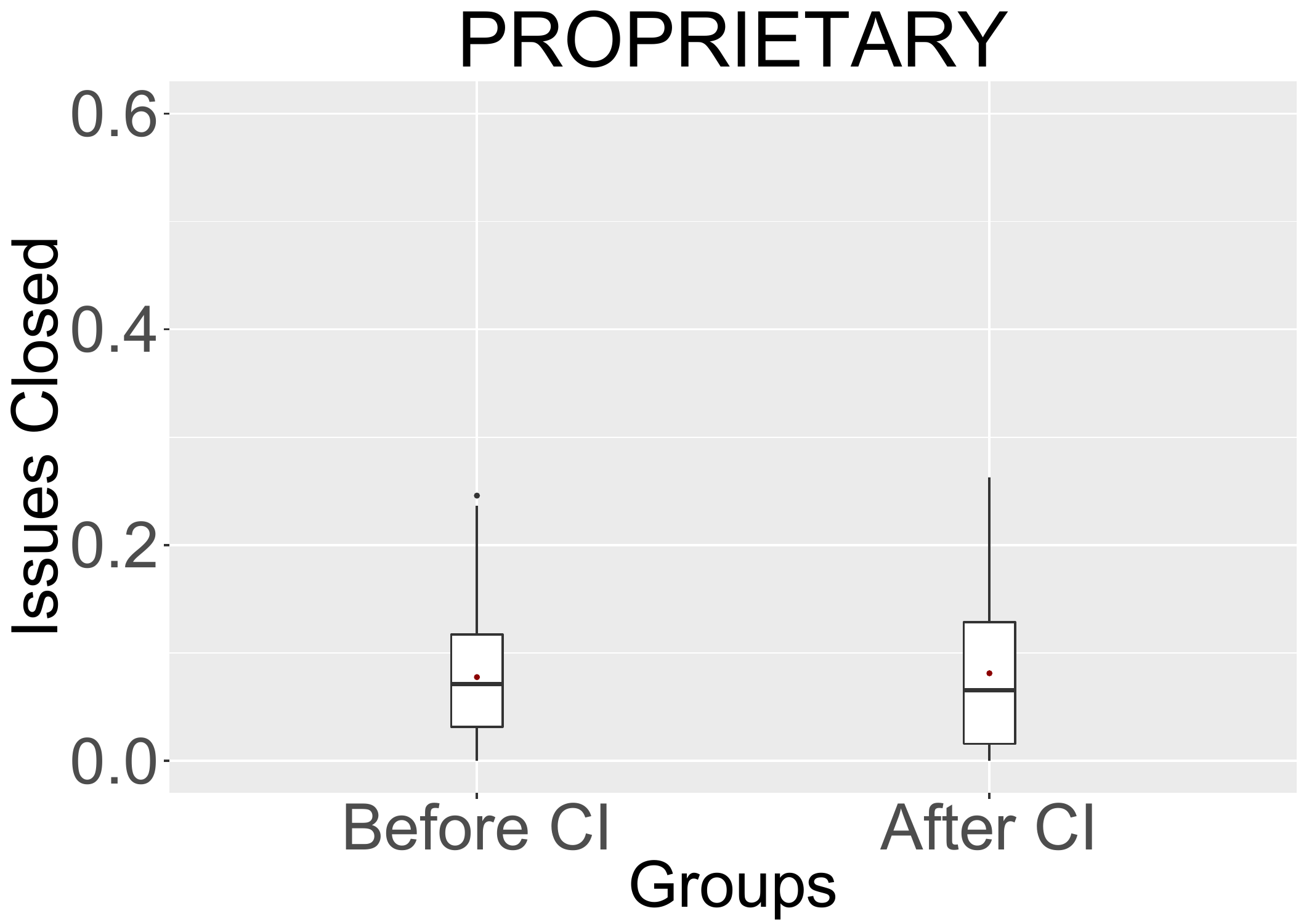}
 \label{fig:res:nci:pro}
}
\subfloat[]{
 \includegraphics[width=0.24\textwidth]{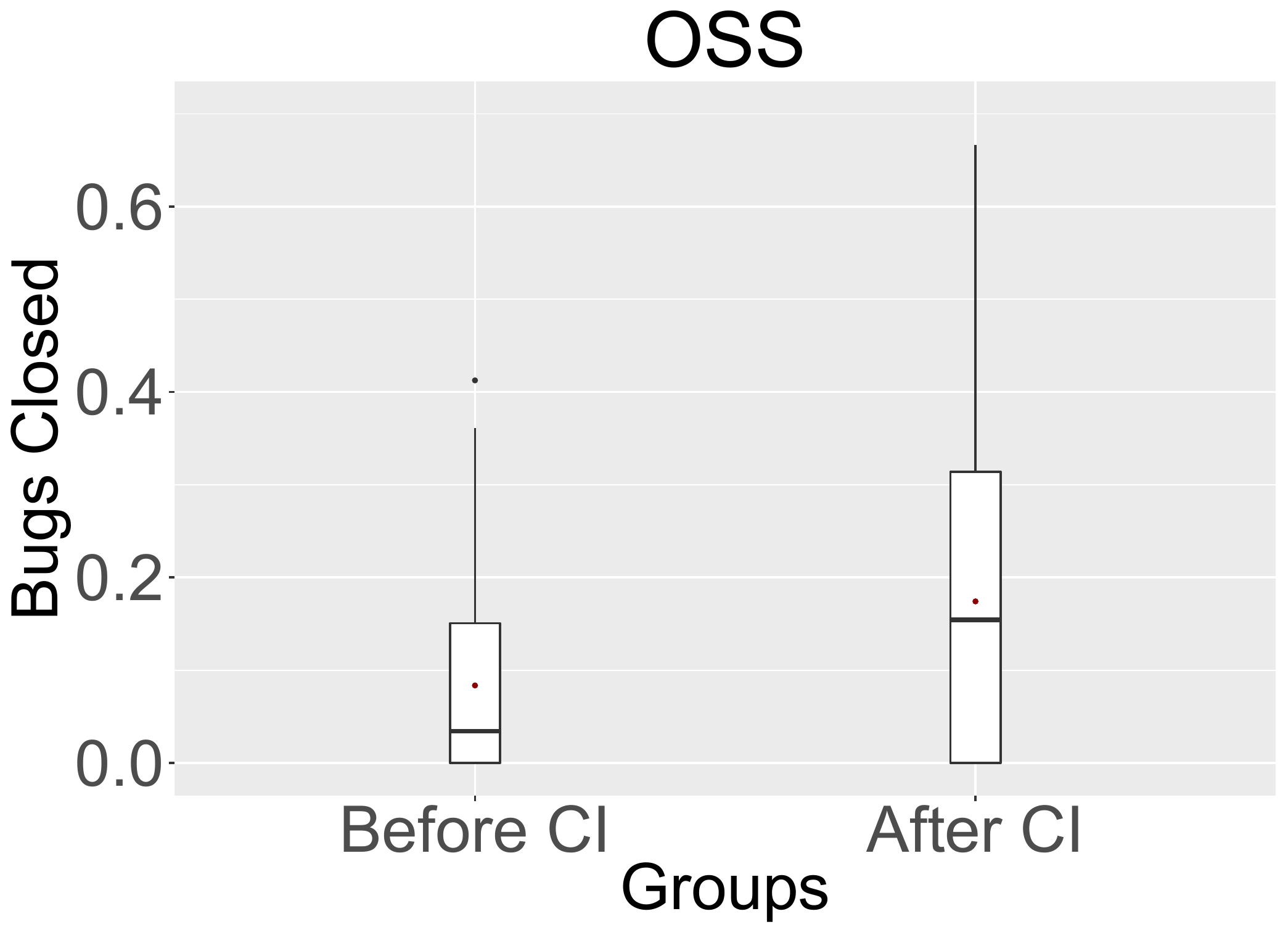}
 \label{fig:res:ncb:oss}
}
\subfloat[]{
 \includegraphics[width=0.24\textwidth]{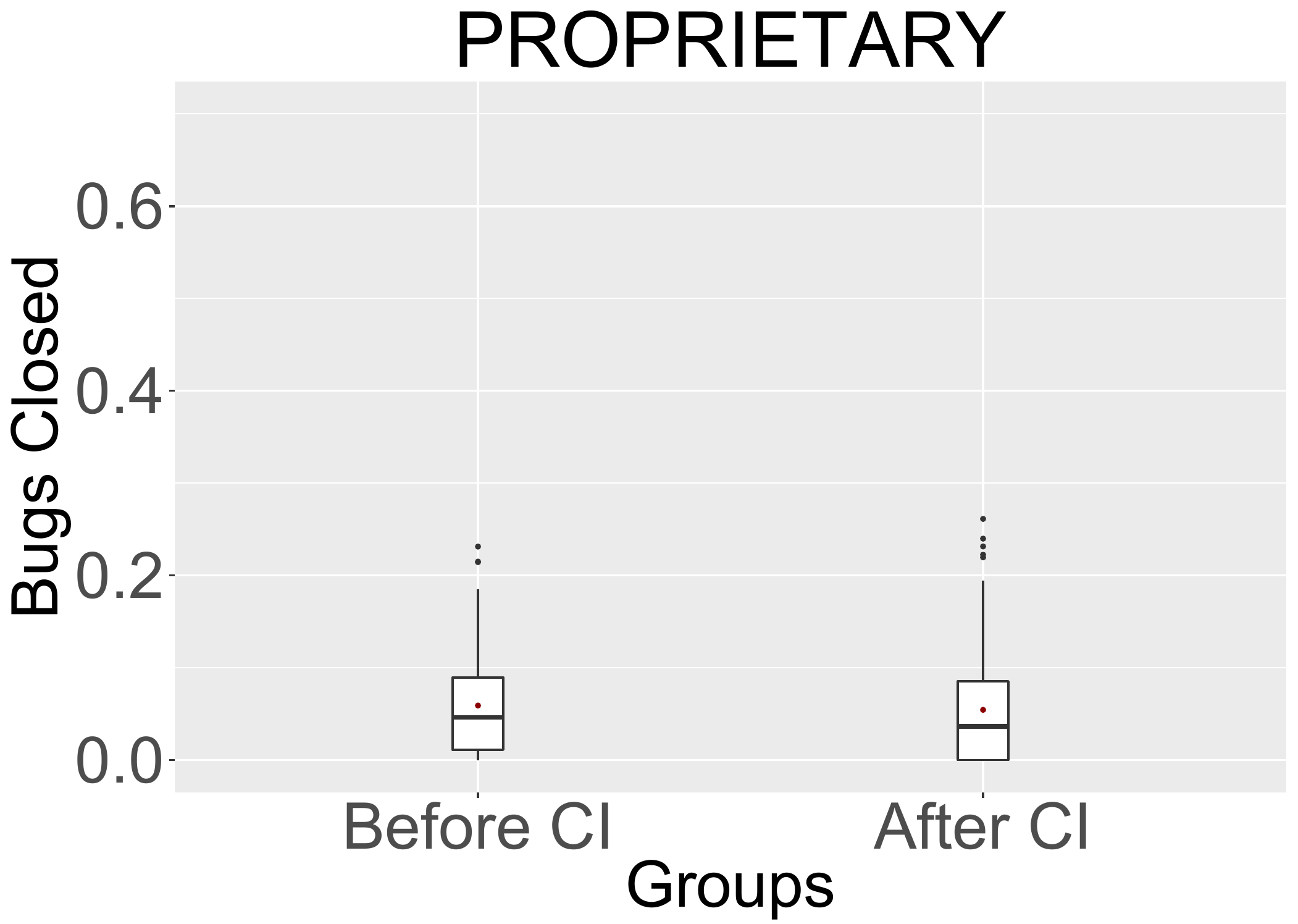}
 \label{fig:res:ncb:pro}
}
\setlength\abovecaptionskip{-3pt}
\setlength\belowcaptionskip{-3pt}
\caption{Normalized closed issues ($NCI$) and normalized closed bugs ($NCB$) for OSS and proprietary projects.} 
\label{fig:res:rq1} 
\end{figure*}

We answer RQ3 by reporting the summary statistics of number of issues that are closed ($NCI$) and number of closed bugs ($NCB$), before and after adoption of CI. In Figures~\ref{fig:res:nci:oss} and~\ref{fig:res:nci:pro}, we respectively report the $NCI$ values for our set of OSS and proprietary projects. 

In Table~\ref{table:rq1}, we report the results of the three statistical measures: the Mann-Whitney U test, effect size, and the `delta' measure. The `delta' value for which we observe no significant difference is highlighted in grey. According to Table~\ref{table:rq1}, for OSS projects, after adoption of CI, significantly more issues are closed ($p-value < 0.001$). On the contrary, for proprietary projects, the influence of CI is not observable for issue resolution. In OSS projects, considering median, the normalized count of closed issues, increases by a factor of 2.4, after adoption of CI, whereas, the normalized count of closed issues almost remains the same for proprietary projects. Our OSS-related findings are consistent with Zhao et al.~\cite{Zhao:ASE17:CI}.

We report the normalized count of closed bugs ($NCB$) in Figures~\ref{fig:res:ncb:oss} and~\ref{fig:res:ncb:pro}, respectively, for our set of OSS and proprietary projects. We report the results of the three statistical measures in Table~\ref{table:rq1}. According to Table~\ref{table:rq1}, for OSS projects, after adoption of CI, significantly more bugs are closed ($p-value < 0.001$). From Figures~\ref{fig:res:ncb:oss} and~\ref{fig:res:ncb:pro} we observe the median $NCB$ to be 0.15 and 0.03, respectively for after and before adoption of CI. Hence, we can state that for OSS projects, bugs are closed five times more after adoption of CI. Similar to issue resolution, our OSS-related findings for bug resolution is somewhat consistent with prior research~\cite{vasilescu2015quality}. We also do not observe CI to influence bug resolution for proprietary projects. 

\noindent
\fbox{\begin{minipage}{26em}
\textbf{\textit{Answer to RQ3: For OSS projects, significantly more normalized issues and bugs are resolved after adoption of CI. For our set of proprietary projects, adoption of CI has no influence on issue and bug resolution.}}
\end{minipage}}

\begin{table}
\centering
\setlength\abovecaptionskip{-2pt}
\caption{Influence of CI on Closed Issues ($NCI$) and Closed Bugs ($NCB$)}
\footnotesize{
\begin{tabular}{ |c|c|c|c|c| } 
\hline
\textbf{Measure}  & $NCI$            &                            & $NCB$             &   \\ 
\hline
                  & \textbf{OSS}     & \textbf{Prop.}             & \textbf{OSS}      & \textbf{Prop.}  \\ 
\hline
Median            & (A:0.31, B:0.13) & (A:0.06, B:0.7)            & (A:0.15, B:0.03)  & (A:0.03, B:0.04)  \\
$\Delta$          & +1.38            & \cellcolor{lightgray}-0.14 & +4.0              & \cellcolor{lightgray}-0.25  \\ 
p-value           & $< 0.001$        & 0.6                        & $< 0.001$         & 0.9 \\ 
Effect size       & 0.5              & 0.0                        & 0.3               & 0.1  \\ 
\hline
\end{tabular} 
}
\label{table:rq1}
\end{table}


\textbf{Summary of the Empirical Study}: We do not observe the expected benefits of CI for proprietary projects. Unlike OSS projects, after adoption of CI, bug and issue resolution does not increase for our set of proprietary projects. Based on our findings, we advise industry practitioners to revise their expectations about the benefits of CI, as only adoption of CI may not be enough to fully reap the benefits of CI.  


\section{Discussion}
\label{discussion}

In this section, we discuss our findings with possible implications: 

\textbf{The Practice of Making Frequent Commits:} Our findings suggest that only adoption of CI tools may not be enough to reap the benefits of CI. As described in Section~\ref{results}, we observe that CI have no influence on bug and issue resolution for proprietary projects. We caution industry practitioners to be wary of the expected benefits from CI adoption, as only adopting and using CI may not be enough to fulfill their expectations. One possible explanation can be attributed to programmers' practice of making less frequent commits which we explain below.  

Standard practice in CI is to use a version control system (e.g., Git). When a programmer makes a commit, the CI tool fetches the code changes, triggers a build that includes inspection checks and/or tests~\cite{Duvall:ci2007}. If the build fails the CI tool provides rapid feedback on which code changes are not passing the inspection checks and/or test cases~\cite{Duvall:ci2007}. In this manner, the CI process provides rapid feedback about code changes to the programmer~\cite{Duvall:ci2007}. The programmer utilizes this feedback to fix the code changes by making more commits, fixing their code changes, eventually leading to more bug fixes and issue completions. Hence, by making more commits, programmers might resolve more bugs and issues. Our explanation related to feedback is congruent with Duvall et al.~\cite{Duvall:ci2007}; they stated ``rapid feedback is at the heart of CI'' and ``without feedback, none of the other aspects of CI is useful''. 

On the contrary to OSS projects, after CI adoption, we have observed that in proprietary projects, change in commit frequency, number of closed bugs, and number of closed issues is non-significant. Based on above-mentioned explanation, we conjecture that for the proprietary projects, programmers are not relying on CI for feedback, and as a result, the commit frequency does not increase significantly, nor does the count of closed bugs and issues. We make the following suggestion: \textit{\textbf{practitioners might be benefited by seeking feedback on submitted code changes from the CI process, by committing frequently}}.

\textbf{Observed Benefits of CI and `Hero Projects':} Another possible explanation can be derived from the `hero' concept observed in proprietary projects by Agrawal et al.~\cite{amrit:icse:seip2018}. They identified projects, where one or few programmers work in silos and do 80\% or more of the total programming effort, as `hero projects'. Agrawal et al.~\cite{amrit:icse:seip2018} reported the prevalence of hero projects amongst proprietary projects, which indicates that regardless of what tool/technique/methodology is being used, majority of the work will be conducted by a few programmers. In case of these projects, even if CI results in increased collaboration, the resolution of bug and issues will still be dependent on the programmers who are doing majority of the work i.e., `hero' programmers. Based on our discussion, we suggest:\textbf{\textit{for proprietary projects the benefits of adopting CI is dependent on what practices practitioners are following, for example, the practice of making frequent commits.}}   

\textbf{Changing Perceptions on CI Adoption:} Practitioners often follow the `diffusion of innovation' rule, which states that practitioners prefer to learn from other practitioners who have already adopted the tool of interest~\cite{akond:agile2015:cd}~\cite{rogers2010diffusion}. Our empirical study can be helpful for practitioners to re-visit their perceptions about CI adoption and use. For example, by reading a success story of CI adoption for an OSS project, a practitioner might be convinced that CI adoption is a good choice for his/her team. In such case, the practitioner's perceptions can be checked and contrasted with empirical evidence. For CI adoption, learning from other practitioners can be a starting point, but practitioners also need to (i) consider their teams' development context factors, and (ii) assess to what extent other practitioners' experiences hold.

\section{Threats to Validity}
\label{threats}

We acknowledge that our results can be influenced by other factors that we did not capture in our empirical study, for example, the prevalence of hero projects. Other limitations of our paper include: 

\textbf{Spurious Correlations}: In any large scale empirical study where multiple factors are explored, some findings are susceptible to spurious correlations. To increase the odds that our findings do not suffer from such correlations, we have: 
\begin{itemize}[leftmargin=*]
\item{applied normalization on metrics that we used to answer our research questions.}
\item{applied two tests: the effect size test and the Mann-Whitney U test to perform statistically sound comparisons. For OSS projects, we compare and contrast our findings with prior research.}
\item{discussed our findings with industry practitioners working for our industrial partner. The practitioners agreed with the general direction of findings: they stated that many teams within their company use a wide range of tools and techniques which does not work optimally for all teams. The practitioners also agreed that there are significant differences between OSS and proprietary software development, and we should not assume these tools and techniques will yield similar benefits. }
\end{itemize}

\textbf{Generalizability}: We acknowledge that the proprietary projects come from our industrial partner. Whether or not our findings are generalizable for other IT organizations remains an open question. We hope to address this limitation in future work.   

\textbf{CI}: In our paper we have adopted a heuristic-driven approach to detect use of CI in a project. We acknowledge that our heuristic is limited tot he three CI tools, and we plan to improve our heuristics by exploring the possibility to add more CI tools. 

\textbf{Bug Resolution}: We have relied on issues marked as a `bug' to count bugs and bug resolution time. In Github, a bug might not be marked in an issue but in commits. We plan to investigate how bugs can inferred from commits, and update our findings accordingly.  



\section{Conclusion}
\label{conclusion}

After mining 150 OSS and 123 proprietary projects, we have quantified the influences of CI on software development for OSS and proprietary projects. We have observed that closed bugs, closed issues, and frequency of commits, significantly increase after adoption of CI for OSS projects, but not for proprietary projects. Our findings suggest that to reap the benefits of CI usage, practitioners should also apply the best practices of CI such as, making frequent commits. We also caution that it may be unwise to hype the usage of CI, promising that CI usage will always increase collaboration, along with bug and issue resolution. While our findings can be biased by our sample of projects, to the best of our knowledge, there exists no large scale research study that reports the opposite of our conclusions. At the very least, our results raise the issue of the benefits of CI tools for proprietary projects--an issue that, we hope, will be addressed by other researchers in future research studies.



\balance
\bibliographystyle{ACM-Reference-Format}

\end{document}